# THE NEED OF TRUSTWORTHY ANNOUNCEMENTS TO ACHIEVE DRIVING COMFORT

Rezvi Shahariar[1] and Chris Phillips[2]

[1]Institute of Information Technology, University of Dhaka, Dhaka, Bangladesh
[2]School of Electronic Engineering and Computer Science, Queen Mary University of London, London, England

*ABSTRACT*

*An Intelligent Transport System (ITS) is more demanding nowadays and it can be achieved through deploying Vehicular Ad Hoc Networks (VANETs). Vehicles and Roadside Units (RSUs) exchange traffic events. Malicious drivers generate false events. Thus, they need to be identified to maintain trustworthy communication. When an authorised user acts maliciously, the security scheme typically fails. However, a trust model can isolate false messages. In this paper, the significance of trustworthy announcements for VANETs is analysed. To this end, a series of experiments is conducted in Veins to illustrate how the trustworthiness of announcements affects travel time. A traffic scenario is created where vehicles detour to an alternate route with an announcement from the leading vehicle. Both true and false announcements are considered. Results confirm that false announcements and refraining from announcements increase travel time. However, the travel time is reduced with trustworthy announcements. From this analysis, it can be concluded that trustworthy announcements facilitate driver comfort.*

## 1. INTRODUCTION

Trust management plays an important role in the successful implementation of an Intelligent Transportation System (ITS). Vehicular communications are prevalent in implementing the ITS to reduce traffic congestion and enable driving comfort for road users. However, vehicular communications are not currently common, but they are starting to appear in traffic systems as roadside infrastructure is increasingly deployed particularly in major cities and highways. For example, in London and England, there are roadside displays which ask drivers to turn off their engines while they are waiting at certain junctions. Also, roadside displays warn road users by showing special types of messages. For example, they might inform of a road accident occurring one mile ahead of the vehicle's current location and recommend drivers to detour to another route. Nowadays, autonomous, and self-driving cars are seen on roads, but they are not prevalent. They are anticipated to be the future mode of transport as they are built using artificial intelligence which can avoid collisions. Also, the cost of manufacturing is expected to reduce allowing them to be affordable for most people.

The announcement of traffic events guides vehicles to detour but sometimes malicious drivers/vehicles announce false messages to deceive other vehicles/drivers. As a result, they detour away from their original route unnecessarily. This may increase travel time as well as reduce their driving comfort, influenced by the malicious sender's message. Alternatively, when there is a traffic jam and vehicles are not warned, they will queue up around the event which exacerbates the traffic congestion. Thus, it is desirable to announce trustworthy messages regarding the true road situation. A false announcement might claim there is an accident when there is none and vice versa. It makes sense to create and manage an environment where only trustworthy messages can flow. A security approach in itself cannot protect the network from these false announcements as they could be generated by malicious authorised users. A strong security approach confirms authentication, integrity, nonrepudiation as well as availability. Trust management is used to prevent false announcements additional to a security scheme. In some cases, both techniques work together to thwart attacks from both internal and external sources. Here, external sources mean attacks initiated from unauthorised vehicles. This is possible because messages are announced in an open wireless medium which are rebroadcast to any vehicle which is in direct range of the relayer or sender. Therefore, trust management can enrich security mechanisms by differentiating malicious vehicles from trusted ones and helps maintain users traffic comfort. The contributions of this paper are as follows:

a) First, a short review is performed on existing trust models which work with false message dissemination in VANETs. Then the necessity of trustworthy announcements is highlighted as it impacts the travel time of vehicles, and which also affects driver comfort in terms of travel time avoiding congestion versus the additional transfer time of the longer route.
b) After this, a generic communication model which follows most existing receiver-side evaluation-based trust models is discussed. These models exhibit higher communication overhead and decision-making time due to the nature of their communication.
c) Next, we focus on a sender-side evaluation of trust to minimize these metrics so that driving comfort can be maintained as vehicles only take the time they truly require to reach their destination.

In this paper, Section 2 introduces a vehicular ad hoc network (VANET), identifies its key elements, and highlights the communication mechanism. Section 3 reviews existing trust models which thwart false messages from spreading in a VANET. In Section 4, the necessity of trustworthy announcements is analysed by measuring the average travel time of all participant vehicles to show less journey time is required when trustworthy messages are announced. Then in Section 5, a generic communication model, which is used by most receiver-side evaluation-based trust models, is illustrated. In Section 6, the criteria for effective trust management are discussed where a sender-side evaluation-based trust model is used as an example. Finally, Section 7 concludes this paper by asserting that more research on sender-side trust schemes is required to enable timely detours whenever available.

## 2. VEHICULAR AD HOC NETWORK COMPONENTS AND COMMUNICATION

A Vehicular Ad Hoc Network (VANET) is a wireless network consisting of vehicles fitted with an On-board Unit (OBU), Roadside Units (RSUs), and an optional Central/Trust Authority (CA/TA). In a VANET vehicles meet each other spontaneously so it is a sporadic and intermittent network. Vehicles move fast, so their topology and distribution change more rapidly than most forms of ad hoc network. In VANETs, vehicles usually move faster (1km/s to 200Km/s) than other wireless nodes where they can exchange event messages within a 0~1000m range. The vehicles/RSUs typically send/receive messages using the IEEE 802.11p-based Dedicated Short Range Communication (DSRC) protocol. A dedicated bandwidth of 5.9 GHz is preserved from the WAVE protocol stack for the successful implementation of VANETs [1]. All main types of vehicles including regular vehicles (private cars), official vehicles (police, ambulance, fire service vehicles) and public authority vehicles (buses, and licensed taxis) can be part of a VANET. This network demands timely traffic updates from the nearby RSUs to avoid problematic regions. Figure 1 depicts a typical VANET where all the key elements are shown.

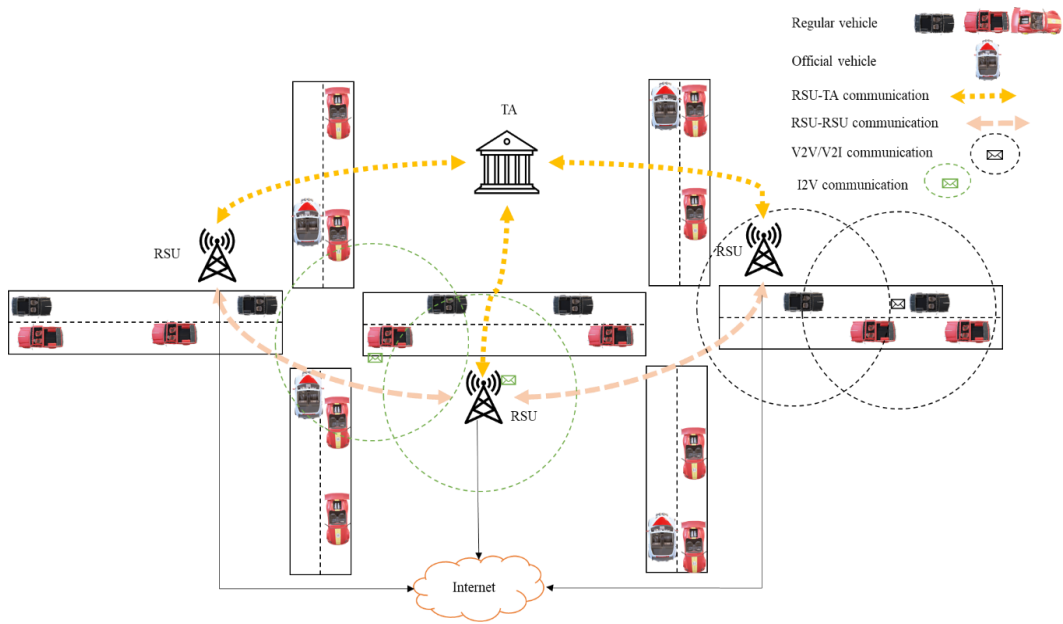

Figure 1. A typical VANET Scenario with its Key Components

Regular vehicles are the primary users of a VANET. They broadcast periodic beacons and announce traffic events when there is one on the road. A sender vehicle is the one which originates the message. Vehicles that receive the event are called receivers. Upon reception, they may act on the event and/or retransmit it to neighbours so that nearby vehicles are informed about the traffic incident. In this way, one vehicle helps neighbouring vehicles from getting stuck in congestion or avoiding an undesirable situation on the road. An intermediate vehicle that forwards a message is called a message relayer. Each vehicle is pre-equipped with a transceiver to send/receive with other OBUs and RSUs [2]. The OBU may be further equipped with an Event Data Recorder (EDR), and Global Positioning System (GPS) sensor [2]. It is also common to utilize a Tamper-Proof Device (TPD) to hold data records or to perform some data manipulation [3, 4]. Many types of official vehicles may be present on roads. Among them, police, ambulance, and fire service vehicles are the most frequent. They visit an event location to intervene when instructed. When the event is resolved, they typically announce a message about its resolution so regular vehicles can use that road again. Event updates from them can be considered completely authentic. RSUs are placed alongside the road to broadcast timely traffic updates to nearby vehicles. RSUs can communicate themselves either using a dedicated broadband network or using a wireless network. Also, they are connected to the Central/Trust Authority (CA/TA) through a dedicated wired or wireless Internet connection. RSUs send information about traffic incidents to the CA/TA.

Vehicles receive periodic traffic updates as well as emergency events from RSUs [2, 5]. Furthermore, RSUs treat messages from official vehicles as "high priority" when they are attending specific emergency events. The TA/CA is the ultimate authority in a VANET. The TA registers vehicles/RSUs, authenticates vehicles, and blacklists malicious vehicles when the extent of their malicious activity exceeds a threshold. It is mandatory to place the TA/CA in a highly secure environment. Furthermore, the TA/CA must be equipped with sufficient computing resources to fulfil the demands of processing requests from other entities. In this network, a vehicle announces a message, which is relayed by intermediate vehicles to reach neighbour vehicles. That means, both single and multiple-hop communication are required. When there is no vehicle in the direct range, messages are simply dropped. This network requires broadcasting of events at the right time, otherwise, traffic congestion, or other undesirable phenomena may emerge on the road. VANET allows vehicles and RSUs to communicate and share news of traffic situations. This way other vehicles can avoid traffic congestion which improves driving comfort and helps them reach their destination on time.

## 3. REVIEW OF TRUST MODELS

Vehicles in VANETs verify the messages and/or sources to determine the reliability of the information. In this way, vehicles identify an untrue message dispersed across the network from a malicious vehicle. This malicious information may cause some vehicles to detour and in some cases, vehicles may queue up around an event which results in traffic congestion or more serious events. Subsequently, this affects the comfort of drivers who believe the untrue information. A trust model and security approach are used to thwart this type of attack. Mostly, schemes run a false message detection method after the arrival of messages. Furthermore, several approaches limit the dissemination of false data in VANETs [6, 7]. While validating traffic events, some methods accumulate recommendations/feedback at RSUs, while others accumulate data at receiver vehicles. When an RSU accumulates recommendations, it disseminates a corrective message after the discovery of malicious data. If vehicles collect recommendations, they individually decide the trustworthiness of an event based on the information they possess. However, as RSUs have wider knowledge and are fixed in the event zone, they can collect more messages than vehicles which leads to more accurate detection of false data than vehicular detection schemes.

In [8], the trust scheme utilizes a false message detection system to produce feedback on the received message which is combined with the reputation to estimate the trust. Vehicles use primary and secondary scores from the RSUs for future communication until the next update. This scheme is analysed using false messages in both urban and highway settings. In [9] the researchers present a Blockchain-Based Traffic Event and Trust Verification (BTEV) framework. This manages trust, privacy, and security using a two-stage verification of events and a two-phase transaction for fast notification of events. The scheme can thwart selfish behaviour and false message rebroadcasting. In [10], the researchers present an infrastructure-based scheme called TRIP which considers the severity level of the safety messages. Moreover, this approach computes the trust of the sender using direct interaction and indirect recommendations. In their evaluation of the model, they generate false messages from malicious vehicles. This model effectively detects malicious action when the malicious vehicle rate is less than 50%. This approach is simple, fast, accurate, scalable, and resilient to some threats. Conversely, the authors in [11] apply fuzzy logic to calculate the trust using experience, plausibility, and location accuracy. This approach determines location accuracy using fog nodes. It can detect false attacks and message alteration attacks.

Reference [12] proposes a trust model to compute trust with high accuracy using a high ratio of malicious vehicles. This model evaluates an event using a coefficient-based weighted mechanism which uses external information, sensed and self-experience (internal information) relating to the event. The final trust is then compared against a limit to determine the validity of an event, for example, false or genuine. Also, a reinforcement model is developed to change the trust evaluation function based on previous results. Vehicles which conform to the protocol are normal and those that do not comply with the protocol either intentionally (malicious) or unintentionally (faulty) are considered malicious. The model is evaluated in Veins using real-world map data and the precision is compared with voting-based, Bayesian, and DST-based approaches by varying the influence of the false information. The researchers in [13] present a model that computes the trustworthiness of the message. This approach considers one-hop Emergency Warning Messages (EWM) and multi-hop Event Reporting Messages (ERM). Receiver vehicles gather messages from in-front vehicles and those vehicles which only pass the event location. Vehicles determine the validity concerning a received message using the location and time closeness of the event and whether it comes from the leading vehicle, or if it is driving through the region later. Then each receiver uses a timer and upon expiration, it compares the sum of all confirming with the sum of non-confirming event reports to trust an event. This approach is effective against bad-mouthing attacks, on-off attacks, and sybil attacks. In [14], a vehicle uses cognition to learn from the environment and develops context around an event to suggest trust. It forms an ontology-based context which links a set of interrelated concepts (for example, vehicle, evaluation, event). This framework considers experience, opinion, and role for the trust evaluation. For outlier detection, time, speed, and distance limits are verified. Besides finding the trust of every report, this model also determines the confidence of the report. The framework is simulated in MATLAB for both rural and urban scenarios and compared against existing frameworks using a confusion matrix. This model also employs sending false messages from malicious vehicles.

After conducting this short survey, it is found that existing literature considers false message announcements. So, trust management which can assure only trustworthy announcements can be used in VANETs. This will lower congestion and waiting time for drivers which improves driving comfort.

## 4. TRUSTWORTHY ANNOUNCEMENTS LIMIT JOURNEY TIME

In this section, the influence of trustworthy announcements on driver/vehicle travel time is analysed. Consider a scenario where an announcement of an event on a road allows drivers to detour to another route if there is one.

On the other hand, without an announcement of this event, vehicles simply queue up around the event as they are driving on the planned route which may exacerbate the situation. This scenario is of interest as no announcement of an event creates problems for other vehicles. When this happens, it create congestion or a jam on the road. Additionally, in the presence of an event when no driver announces it, this may also create congestion. On the other hand, a false announcement causes drivers to detour unnecessarily which increases their travel time. Thus, a pair of experiments are conducted in a simulated environment to show that trustworthy announcements can improve driver comfort by maintaining low journey time which allows vehicles to reach their destination in a timely manner. To this end, a road network is constructed in SUMO that has two alternate routes where a diversion message announcement from the first route causes vehicles to detour to the second alternate route.

Every simulation for these experiments runs for 300 seconds as within this period all vehicles can finish their journey. During the simulation, vehicles are inserted periodically one after another using a car following model where each vehicle is inserted with a 1 second insertion gap. Vehicles start their journey from the upper left corner and travel along the primary or alternate route to finish their journey in the upper left corner of Figure 2. After that, the journey time of each vehicle is recorded in all situations, for example, with or without an event. A road situation called a diversion is imagined when a part of the route needs maintenance or maybe there is an emergency event at this location. Vehicles then cannot use the problematic road. To simulate this, we introduce the "halting" of a front vehicle for 120 seconds duration and when the duration expires it announces a diversion message towards others. The vehicles which do not already enter the primary route detour to the alternate route upon the diversion message arrival as shown in Figure 2. As a result, vehicles face increased travel time. This extra driving time attracts attention which can be reduced or can be kept in limit by announcing a timely traffic update from a front vehicle around the event. In these experiments, the journey time of all vehicles is collected in all situations. The following average travel time is measured from the recorded data in the following settings:

a) average travel time of all vehicles when there is no event.
b) average travel time of all vehicles when a vehicle sends a false announcement.
c) average travel time of all vehicles when no one announces a road traffic event.
d) average travel time of all vehicles when a trustworthy announcement takes place.

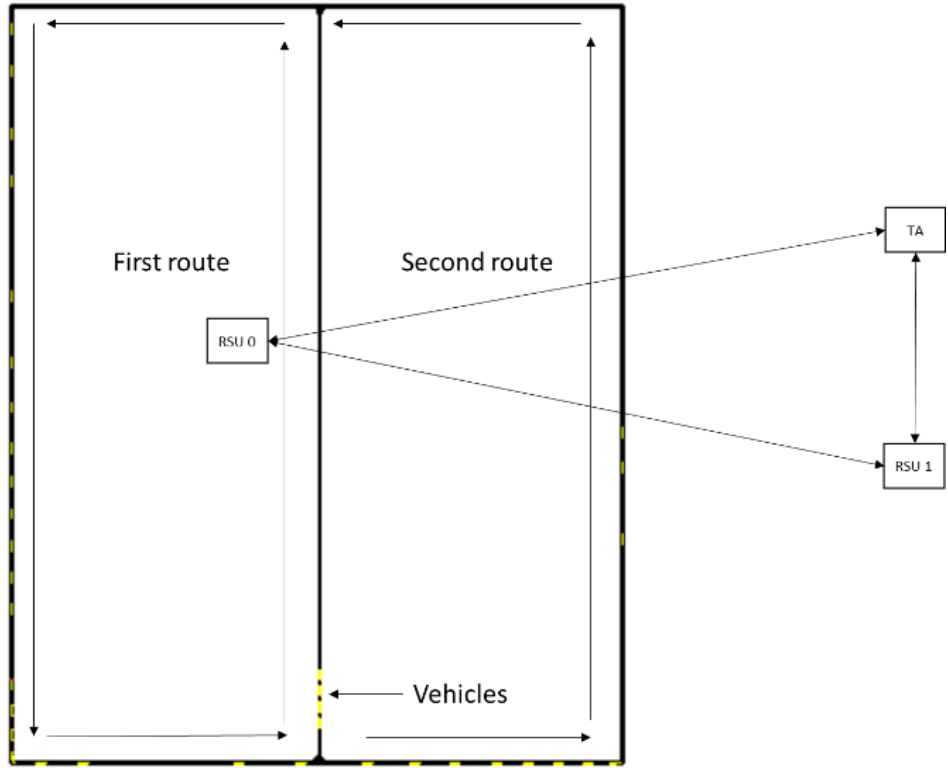

Figure 2. A Road Network with Two Alternate Routes

During the simulation, vehicles normally use the primary route. Vehicles can also detour to a second alternate route when they receive an announcement about an event on the primary route. In these experiments, only two RSUs and one TA are used, although the roles of TA are not the focus of this paper. The experiments are conducted in the presence of 10, 30, and 50 vehicles. First, the average travel time of all vehicles on the primary route is shown when there is no traffic event which is denoted by the blue-coloured line in Figure 3. Then they are compared with the average travel time of all vehicles when an untrue message is announced by vehicle V0 which is denoted by the orange-coloured line in Figure 3. It is clear from Figure 3 that the average travel time of all vehicles is increased by at least 20 seconds in all cases as vehicles are detoured onto the longer second route due to the untrue announcement. This does not happen always when the vehicle density is high, and they are inserted periodically. Some vehicles which are inserted later find the primary route free and use it. When the density is 30, most of the vehicles are near the junction, so they make the detour upon the message's arrival. With 50 vehicles, some vehicles are at the junction, so they take the instant detour. However, other vehicles which arrive later at the junction, and find the primary route free, so they do not detour to the alternate route. This is why the orange-coloured line shows a small decline in average travel time with 50 vehicles relative to the case when the experiment is conducted with 30 vehicles.

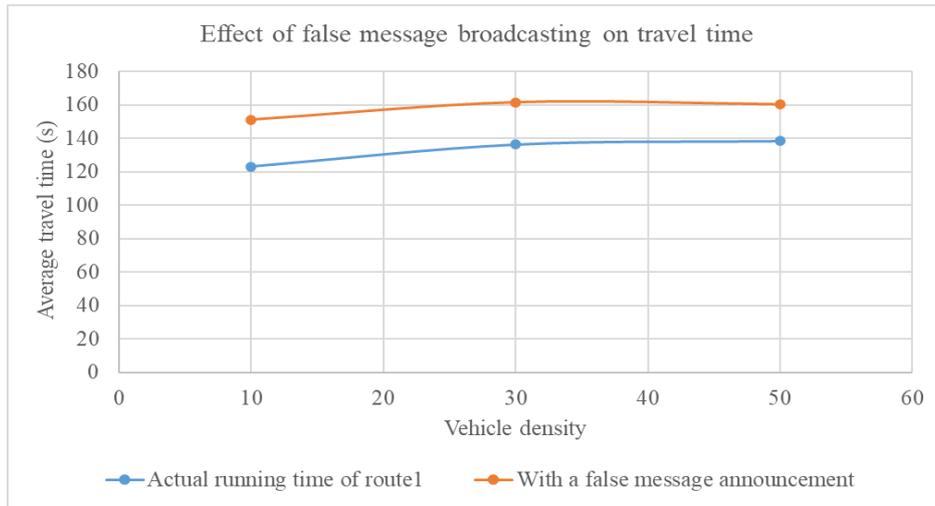

Figure 3. Effect of Untrue Event Announcement on Travel Time

In the second experiment, an unannounced true event on the primary route causes the vehicles to be queued for 120 seconds. This is why, their average travel time is increased by at least 90 seconds more than the normal case in Figure 3. This is identified by the blue-coloured line in Figure 4. However, these values are reduced by at least 36 seconds with a trustworthy message announcement from the V0 as shown by the orange-coloured line in Figure 4. This reduced travel time directly relates to driving comfort in terms of the total journey time of vehicles. In this case, driving comfort is achieved through reducing the journey time of vehicles from trustworthy announcements. For vehicles that do not receive the event announcement, their journey time remains the same in both cases. As the results suggest, improved travel time with trustworthy announcements can be pivotal in improving driver comfort by avoiding delays.

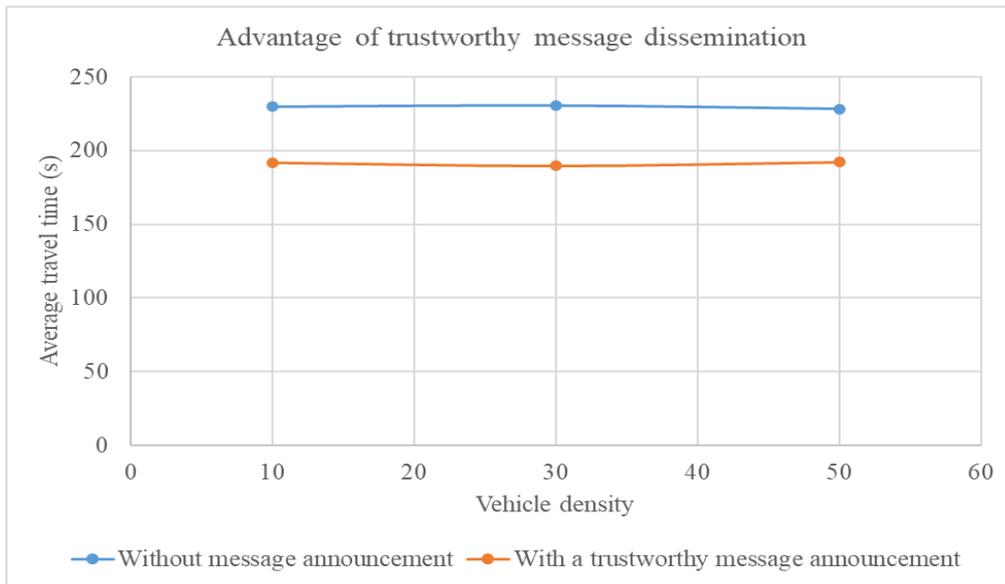

Figure 4. Improved Travel Time with True Event Announcement

# 5. HOW MOST EXISTING MODELS MANAGE TRUST

Whenever a vehicle encounters an event on a road it informs neighbouring vehicles and RSUs. To the best of our knowledge, the receivers in most existing models start computing trust after receiving a message. They do not verify the trust of the sender prior to it announcing a message. As there is no limit on message generation, even an unknown vehicle may broadcast a false emergency message and deceive other vehicles. Figure 5 shows that a message from vehicle V causes vehicles A, B, C, W, X, Y, and Z to compute the trust individually irrespective of whether V is a trusted or untrusted source. Also, they may individually contact each other to obtain direct and indirect trust metrics as well as optionally contacting a nearby RSU to obtain information about the sender vehicle. This may involve further data collection at an RSU to provide feedback to the asking vehicles. After this, RSUs apply an evaluation algorithm to compute the trust of the sender vehicle to decide on the acceptance or rejection of the message.

Figure 6 illustrates the basic decision-making mechanism followed by the trust-based approaches listed in [8, 15-20]. Here, every receiver individually evaluates the trust of every received message and/or the sender. Besides this, they communicate with the RSU and other neighbours to obtain trust information at run time. The RSU also collects data from official vehicles or trusted vehicles about an event. Alternatively, some approaches do not use any infrastructure [21, 22], but they generate a sizeable number of messages among neighbours periodically to manage the trust. For example, the trust approach in [22] uses Bayesian statistics to calculate direct trust and uses Dempster-Shafer theory to determine recommendation trust. Alternatively, the trust model listed in [21] continuously sends and receives hello messages to collect data as well as exchange trust metrics between neighbours. To some extent, these two approaches follow the pattern of interaction shown in Figure 7. The diagram in Figure 8 shows the general sequence of communications in most existing schemes between the actors: RSU, sender, and receiver vehicles. This sequence diagram contains only three types of actors as most models deal with communication among these three entities.

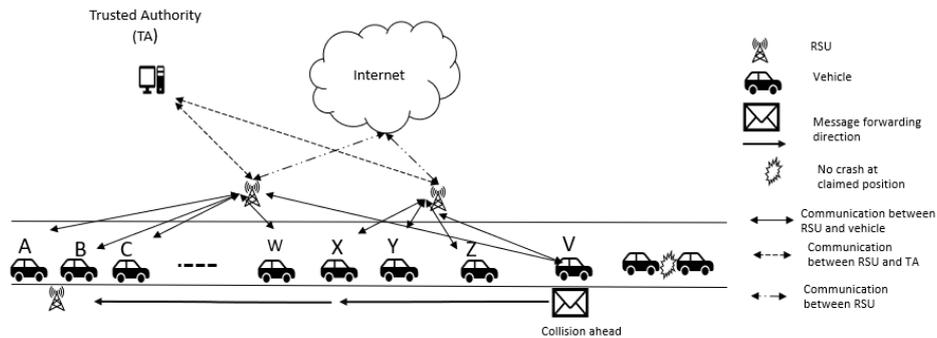

Figure 5. Illustration of an Untrue Message Triggering Trust Computation at All Receivers

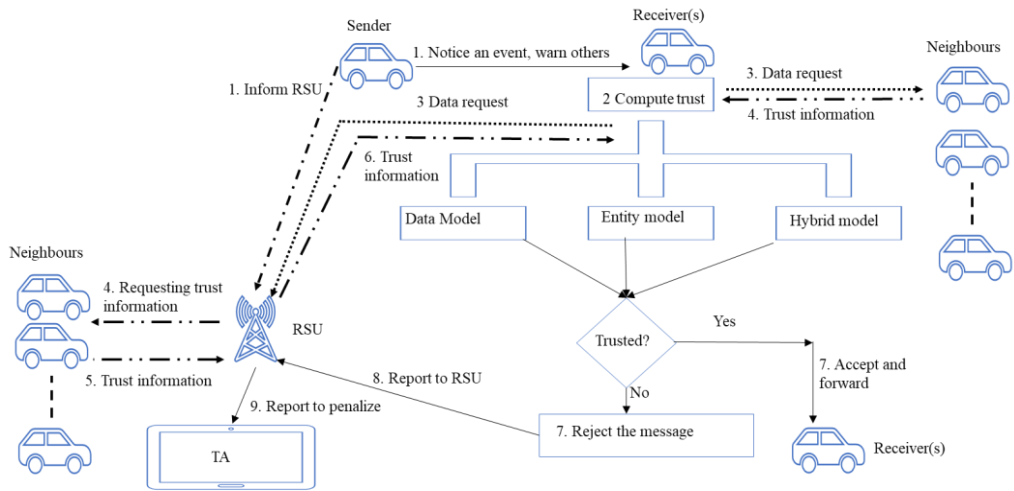

Figure 6. A Generic Model for Trust Verification in Most Receiver End-Based Approaches

1. V can receive direct trust from its direct neighbours U, and W.
2. Additionally, V can collect periodic trust information from indirect neighbours like T, X, and Y.
3. Then, V finds the local trust value for a sender using both direct trust and indirect recommendation information.

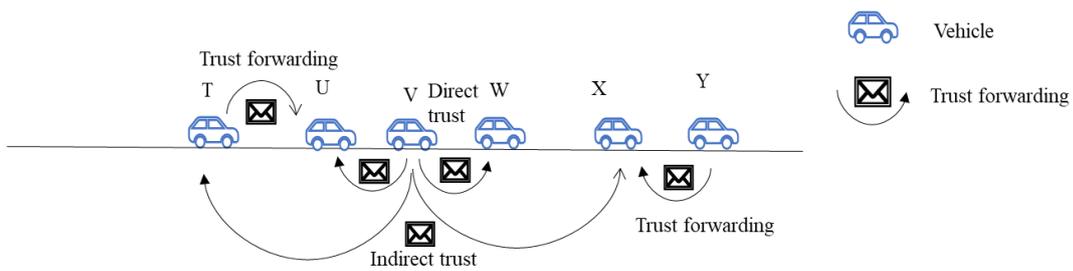

Figure 7. Trust Calculation Using Direct and Indirect Trust

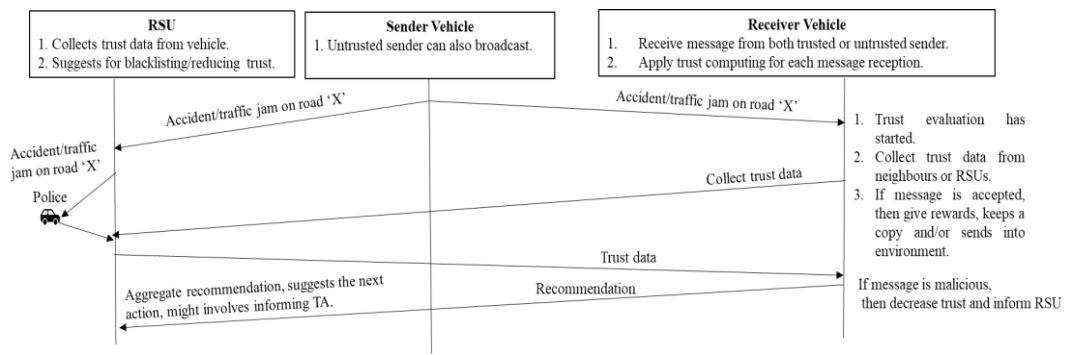

Figure 8. A Sequence Diagram of Event Broadcasting in Existing Trust Models

It is seen in existing models that any vehicle can broadcast messages in VANETs as their trust is not considered at the time of sending messages, as is the case for vehicle V in Figure 5. Thus, V cannot be considered a reliable source until its trust is evaluated at the receiver vehicles. This situation arises in most existing approaches. These approaches discard messages from an untrusted sender only after evaluation which consumes lots of resource. Otherwise, they accept a

21

message and treat the sender as trusted. These schemes result in additional overhead in the network and delay the trust verification process which creates a performance bottleneck. A dedicated bandwidth of 5.9 GHz is reserved with the WAVE protocol stack to support the communication in VANETs. To address the overhead problem, more bandwidth can be reserved but this needs the allocation of more bandwidth from the IEEE 802.11p protocol. In future, we may expect more bandwidth will be reserved for communication. Even so, we should not waste precious bandwidth when managing a single traffic event as we do not know if multiple concurrent traffic events may occur and need to be managed in a region. Additionally, a false recommendation from an indirect source contaminates the resultant trust computation. Indirect recommendations are also a concern if relayed via an untrusted vehicle. To identify fake recommendations, some approaches use additional filtering methods, but this adds an extra level of complexity [22]. Furthermore, when vehicles take more time to decide on an event they may enter a hazardous zone resulting in more traffic turmoil than the original reported event. For example, severe traffic congestion could be seen around an accident if the scheme lacks a fast driver decision time even if the accident event is announced in a timely manner. Thus, the delay in decision causes vehicles to enter the hazardous zone. This case is common with the receiver-side-based evaluation schemes as they evaluate trustworthiness after message arrivals which requires further communication with neighbouring RSUs and vehicles as well as computation. Thus receiver-side trust schemes suffer from higher driver decision times which can lead to more traffic chaos.

There may be some occasions when a vehicle is authorized, but its trust is not yet established. A false message from it abuses the network resources. Also, some approaches either collect trust information from the neighbourhood or globally. However, they do not cope well with rapid topological changes. Other approaches do not cater for high-priority messages from official vehicles. Consider the situation when a vehicle receives an accident announcement from another vehicle. It must decide before moving towards the particular direction. For example, it can select to avoid the area without waiting for the trust evaluation decision, or it can initiate the trust evaluation of the sender and/or its message. The vehicle may then decide on which way it will drive. It may use the original route, or it can select an alternative path. If it acts without awaiting the trust decision, for a false message, it helps the wrongdoer to achieve his/her objective. In most existing approaches, vehicles select the second option which requires time to make a decision. By this point, some vehicles may have entered the affected area. This diminishes the impact of an emergency announcement. The delay until a decision is reached is a concern with existing trust schemes. The slow response time can aggravate a situation.

## 6. TOWARDS EFFECTIVE TRUST MANAGEMENT FOR VANETS

Sender-side trust evaluation is a comparatively new research direction for trust evaluation in VANETs which would benefit from more attention from researchers. This is an interesting evaluation scheme as it allows receivers to believe messages immediately. Sender-side schemes reduce the demands on network resources and latency when an event occurs. Moreover, messages from an untrusted source should not be announced. It is therefore better for a network to restrict messages from an untrusted source. With suitable access control, many independent evaluations at receiver vehicles can be avoided as the likelihood of sending untrue messages from an untrusted sender should be suppressed. When VANETs employ sender-side trust evaluation, receiver vehicles are free from the verification of the source vehicle and/or its message. Also, receivers do not need to communicate with further vehicles and RSUs to obtain updated trust values. Hence, they can decide early about the truthfulness of the event, or they can send an attack reporting message towards other vehicles and RSUs to verify the original sender's message if there is doubt. This verification will be conducted only by a responsible entity like an RSU by collecting some feedback from the trusted vehicles which visit the event location. After this, the RSU can disseminate a corrective message towards the vehicles which are visiting the event location.

This system allows low communication overhead as it avoids trust evaluation of the sent messages in most cases. Only in some exceptional cases further verification of the original message may be needed, whenever there is cause to refute a message from an original sender vehicle, or if a receiver is suspected of sending a false report of the original message. When this happens, the mischievous driver receives a punishment which lowers its trust resulting in subsequent constrained announcements or access blocking in extreme cases. However, false reporting of events only happens occasionally as there are punishments for misbehaving. Thus, this system allows lower communication overhead compared to receiver-side evaluation and allows for faster driver decision times as drivers do not generally need to verify messages. This model maintains an environment where only trusted drivers can announce messages and receivers do not need to wait for verification. Consequently, vehicles can immediately detour when there is an alternate route to use and avoid the hazardous zone. Using sender-side control of the trust score, receiver vehicles are no longer required to wait for any decision from other trust entities and can instantly decide on the appropriate action. In [23], the authors presented an approach using this concept to minimize communication overhead and response time. This trust model incorporates a fuzzy logic-based reward and punishment assessment when there is a dispute between the sender and reporter(s) [24]. The parameters for fuzzy judgement are the driver past behaviour, severity of incident, and RSU confidence of the sender/reporter. In future, more focus is needed on the sender-side evaluation techniques to enrich the trust model with novel ideas and features. Consequently, in the presence of RSUs along the roads to monitor and control this network, this trust model promotes an environment where only trustworthy announcements will be disseminated to reduce driver delays and frustration.

## 7. CONCLUSION

Overall, in this paper, the necessity of trust management is illustrated using a series of simulations conducted in Veins which suggests that trustworthy announcements can maintain lower travel times for drivers. This in turn relates to driver comfort as vehicles can detour in the presence of an event which helps to avoid congestion and frustration. Then, two forms of generic communication models seen in most existing trust models are presented. These trust models follow receiver-side evaluations to verify messages when they arrive at receivers. It is found that all receivers independently or cooperatively determine the trust of events. Hence, they tend to exhibit higher communication overhead and driver decision times compared to sender-side evaluation of trust. Sender-side evaluation-based trust does not require verifying the trust of every received message, so it does not require communicating with other vehicles or RSUs for trust metrics exchanges which improves efficiency and driver comfort in VANETs.

## AUTHORS

**Rezvi Shahariar** received his B.Sc. degree in Computer Science from the University of Dhaka, Bangladesh in 2006; and an M.S. degree in Computer Science in 2007 from the same institution. After some time as a Lecturer at the University of Asia Pacific, Dhaka, Bangladesh, he is now an Assistant Professor at the Institute of Information Technology, University of Dhaka. Then he obtained a PhD on trust management framework for Vehicular Ad Hoc Networks (VANETs) from Queen Mary, University of London (QMUL). His research interests include wireless network analysis with an emphasis on trust, security in VANETs, and the application of machine learning to security.

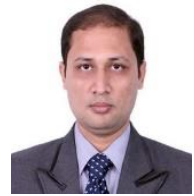

**Chris Phillips** (MIEEE) received a BEng. Degree in Telecoms Engineering from Queen Mary, University of London (QMUL) in 1987 followed by a PhD on concurrent discrete event-driven simulation, also from QMUL. He then worked in industry as a hardware and systems engineer with Bell Northern Research, Siemens Roke Manor Research and Nortel Networks, focusing on broadband network protocols, resource management and resilience. In 2000 he returned to QMUL as a Reader. His research focuses on management mechanisms to enable limited resources to be used effectively in uncertain environments.

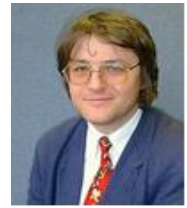